\documentclass[
aps,%
12pt,%
final,%
notitlepage,%
oneside,%
onecolumn,%
nobibnotes,%
nofootinbib,
superscriptaddress,%
noshowpacs]%
{revtex4}

\usepackage[dvips]{graphicx}

\newcommand{\egg}{\lowercase{\eta^\prime g^* g^*}}
\newcommand{\eggr}{\lowercase{\eta^\prime g^* g}}

\begin{document}

\begin{flushright}
DESY 01-012 \\
YARU-HE-01/01 \\
February 2001\\
\end{flushright}

\vspace*{1.5cm}
\begin{center}
{\Large \bf
\centerline{The Effective $\eta^\prime g^* g^*$ Vertex
       at Arbitrary Gluon Virtualities}}
\vspace*{1.5cm}
{\large A.~Ali}
\vskip0.2cm  
Deutsches Elektronen-Synchrotron DESY, Hamburg \\
Notkestra\ss e 85, D-22603 Hamburg, FRG\\
\vskip0.5cm
{\large A.Ya. Parkhomenko}
\vskip0.2cm
Department of Theoretical Physics, Yaroslavl State University,
Sovietskaya 14,\\ 150000 Yaroslavl, Russia

\vspace*{6.0cm}

Invited Talk; Conference of the Nuclear Physics Department of the
Russian Academy of Sciences: Fundamental Interactions of Elementary
Particles,\\ Moscow, Russia, November 27 -- December 1, 2000.

\end{center}
\thispagestyle{empty}

\newpage   
\setcounter{page}{1}
\title{The Effective $\eta^\prime g^* g^*$ Vertex 
       at Arbitrary Gluon Virtualities}%

\author{\firstname{A.} \surname{Ali}}
\email{ali@x4u2.desy.de}
\affiliation{%
DESY, Hamburg, Germany
}%

\author{\firstname{A.Ya.} \surname{Parkhomenko}}
\email{parkh@uniyar.ac.ru}
\affiliation{%
Yaroslavl State University, Yaroslavl, Russia
}%


\begin{abstract}
\bigskip
The effective $\eta^\prime g^* g^*$ vertex is studied in the 
standard (Brodsky-Lepage) and modified hard scattering approaches for
arbitrary gluon virtualities in the time-like and space-like regions. 
The contribution of the gluons in the $\eta^\prime$-meson is
taken into account, and the wave-function is constrained
using data on the electromagnetic transition form factor of the
$\eta^\prime$ meson. Our results have implications for
the inclusive decay $B \to \eta^\prime X$ and exclusive decays, such as
$B \to \eta^\prime (K, K^*)$, and in hadronic production processes
$N + N (\bar N) \to \eta^\prime X$. 
\end{abstract}

\maketitle

\section{Introduction}
\label{sec:introd}

The effective coupling involving two gluons and the $\eta^\prime$-meson
enters in a number of production and decay processes. For example, the
inclusive
decay $B \to \eta^\prime X_s$~\cite{cleo-etap-incl} and the exclusive
decay $B \to \eta^\prime K$~\cite{cleo-etap-excl,babar-etap}, involve,
apart from the matrix elements of the four-quark operators, the transitions
$b \to s g^*$, followed by $g^* \to \eta^\prime g$~\cite{TFF-3,TFF-2},
$b \to s g g $ followed by $g g \to \eta^\prime$~\cite{ACGK98},
as well as the transitions $g^* g^* \to \eta^\prime$ and
$g^* g \to \eta^\prime$~\cite{TFF-1}. Thus, a reliable determination
of the vertex function, which is often called the transition form factor, 
$F(q_1^2, q_2^2, m_{\eta^\prime}^2)$ (here, $q_1^2$ and $q_2^2$ represent
the virtualities of the two gluons) is an essential input in a quantitative
understanding of these and related decays. Apart from the mentioned
$B$-decays, the $\eta^\prime g^* g^{(*)}$ vertex plays a role in a large
number of processes, among them are the radiative decay
$J/\psi \to \eta^\prime \gamma$ and the hadronic production processes
$N + N (\bar N) \to \eta^\prime + X$, where $N$ is a nucleon.
The QCD axial anomaly~\cite{anomaly}, responsible for the bulk
of the $\eta^\prime$ mass, normalizes the vertex function on the gluon
mass-shell, yielding $F(0, 0, m_{\eta^\prime}^2)$. The question that still
remains concerns the determination of the vertex for arbitrary time-like
and space-like gluon virtualities, $q_i^2$; $i=1, 2$.
A related aspect is to understand the relation between the
$\eta^\prime g^* g^*$ vertex and the wave-function of the
$\eta^\prime$-meson. Stated differently, issues such as the transverse
momenta of the partons in the $\eta^\prime$-meson and their impact on the
$\eta^\prime g^* g^*$ vertex have to be studied quantitatively.

  While information on the $\eta^\prime g^* g^*$ vertex is
at present both indirect and scarce, its electromagnetic counterpart
involving the coupling of two photons and the $\eta^\prime$-meson,
$F_{\eta^\prime \gamma^* \gamma}$, more generally the meson-photon
transition form factor, has been the subject of intense theoretical
and experimental activity. In particular, the hard scattering
approach to transition form factors, developed by Brodsky and
Lepage~\cite{BL}, has been extensively used in studying
perturbative QCD effects and in making detailed comparison with
data~\cite{BL-Review}. 
A variation of the hard scattering approach, in which transverse degrees   
of freedom are included in the form of Sudakov effects in transition form
factors~\cite{Collins,BS89}, has also been employed in data analyses.
For a critical review and comparison of the standard
(Brodsky-Lepage) and modified hard scattering (mHSA) approaches, see
Refs.~\cite{MR97,Stefanis99}. We note that either of these 
approaches combined with data~\cite{eta'-gamma} 
constrains the input wave-function for the quark-antiquark part of the
$\eta^\prime$-meson. However, the gluonic part of the $\eta^\prime$-meson
wave-function is not directly measured in these experiments and
will be better constrained in future experiments sensitive to 
the $\eta^\prime g^* g^*$ vertex.

We have used the hard scattering approach to study the $\eta^\prime g^*
g^*$ vertex, incorporating the information on the wave-function and the
mixing parameters entering in the $\eta - \eta^\prime$ complex from
existing data involving the electromagnetic transitions.
The results and details of the calculation have been presented by us in
Ref.~\cite{AP2000}. Prior to our work, Muta and Yang~\cite{Muta}
derived the $\eta^\prime g^* g$ vertex, with one off-shell gluon in the
time-like region, in terms of
the quark-antiquark and gluonic parts of the $\eta^\prime$ wave-function,
taking into account the evolution equations obeyed by these partonic
components. In Ref.~\cite{AP2000}, we also addressed the same
issues along very similar lines. We first rederived the $\eta^\prime g^*
g$ vertex, pointing out the agreement and differences between our results
and the ones in Ref.~\cite{Muta}. The differences have to do with the
derivation of the
leading order perturbative contribution to the gluonic part of the
$\eta^\prime g^* g$ vertex, and the use by Muta and Yang \cite{Muta} of
the anomalous dimensions derived in Ref.~\cite{Ohrndorf:1981uz} in the
evolution of the wave-functions. For the anomalous dimensions, we use
instead  the results derived  in Refs.~\cite{Shifman:1981dk,Baier:1981pm},
which are at variance with the ones given in Ref.~\cite{Ohrndorf:1981uz},
but which have been  recently confirmed by Belitsky and
M\"uller~\cite{Belitsky}. Making use of the $\eta^\prime \gamma^* \gamma$
data to constrain the $\eta^\prime$-meson wave-function parameters, we
find that the gluonic contribution in the $\eta^\prime$-meson is very
significant. We then extended our analysis to the case when
both the gluons are virtual, having either the time-like or space-like
virtualities. We studied the effects of the transverse momentum
distribution involving the constituents of the $\eta^\prime$-meson and
took into account soft-gluon emission from the partons by
including the QCD Sudakov factor, following techniques which were 
introduced in studies of the electromagnetic and transition form factors
of the
mesons~\cite{JKR96,JK93,Stefanis99}. We showed the improvements in the
$\eta^\prime g^* g^{(*)}$ vertex function due to the inclusion of the
transverse-momentum and Sudakov effects, which is particularly marked
in the space-like region, improving the applicability of the hard
scattering approach to lower values of~$Q^2$. These effects have a 
bearing on the hard scattering approach to exclusive non-leptonic
decays~\cite{BBNS99}; the importance of the transverse-momentum and
Sudakov effects in the decays $B \to \pi \pi$ and $B \to K \pi$ has also
been recently emphasized in~\cite{KL00}. We also derived
approximate formulae for the $\eta^\prime g^* g^{(*)}$ vertex
\cite{AP2000}, which satisfy the axial-vector anomaly result for on-shell
gluons and have the asymptotic behavior in the large-$Q^2$ domain,
determined by perturbative QCD.

  We summarize in this report the input, salient features of the
derivation of the $\eta^\prime g^* g^{(*)}$ vertex function, and some
numerical results presented by us in Ref.~\cite{AP2000}, making comparison
with the earlier work on this subject.

\section{Definition of the $\egg$ Vertex}
\label{sec:egg-def} 

The $\eta^\prime$-meson, not being an SU(3)$_F$ flavor-octet meson state,
has a gluonic admixture in its wave-function in addition to the usual
quark-antiquark content. We take the parton Fock-state
decomposition of the $\eta^\prime$-meson wave-function as follows:  
\begin{equation} 
| \eta^\prime > \, = \sin \phi \, | \eta^\prime_q > + \cos \phi \, |
\eta^\prime_s > 
+ \, | \eta^\prime_g > ,  
\label{eq:WF-decomp}
\end{equation}
where the SU(3)$_F$ symmetry among the light $u$, $d$ and $s$~quarks 
is assumed. Thus,
$| \eta^\prime_q > \, \sim | \bar u u + \bar d d > / \sqrt 2$ and
$| \eta^\prime_s > \, \sim | \bar s s >$ are the quark-antiquark Fock
states, where $\phi = 39.3^\circ \pm 1.0^\circ$ is the mixing
 angle~\cite{FKS98}, and
$| \eta^\prime_g > \, \sim | g g >$ is the two-gluon Fock state.
All other higher Fock states are ignored.
Following Refs.~\cite{Ohrndorf:1981uz,Shifman:1981dk,Baier:1981pm}, we
express the wave function of the
$\eta^\prime$-meson in terms of the eigenfunctions of the
quark-antiquark $| \bar q q >$ and gluonic $| gg >$ states:  
\begin{eqnarray} 
\Psi & = & C \left [ \phi^{(q)} (x,Q) + \phi^{(g)} (x,Q) \right ] , 
\label{eq:WF-tot} \\
C & = & \sqrt 2 \, f_q \sin \phi + f_s \cos \phi ~,   
\label{eq:C-const}
\end{eqnarray} 
where $f_q$ and $f_s$ are the decay constants
of $|{\eta^\prime}_q>$ and $|{\eta^\prime}_s>$, respectively, 
and their present estimates are: $f_q = (1.07\pm 0.02) \, f_\pi$,
$f_s = (1.34 \pm 0.06) \, f_\pi$, where $f_\pi \simeq 131$~MeV
is the pion decay constant~\cite{FKS98}. The
eigenfucntions are then  calculated for a given $Q^2$ by solving the
evolution equations. As is well known, the results for both the
quark-antiquark 
and gluonic components can be presented as infinite series involving 
Gegenbauer
polynomials~\cite{Ohrndorf:1981uz,Shifman:1981dk,Baier:1981pm}. Including
the leading and next-to-leading terms
in the expansion, the functions  $\phi^{(q)} (x,Q)$ and $\phi^{(g)}
(x,Q)$ are~\cite{AP2000}:   
\begin{eqnarray}
\hspace*{-15mm} &&
\phi^{(q)} (x, Q) = 6 x \bar x
\left \{ 1 +
\left [ 6 B^{(q)}_2
\left ( \frac{\alpha_s (Q^2)}{\alpha_s (\mu_0^2)} \right )^{\frac{48}{81}}
- \frac{B^{(g)}_2}{15}
\left ( \frac{\alpha_s (Q^2)}{\alpha_s (\mu_0^2)} \right
)^{\frac{101}{81}}
\right ] (1 - 5 x \bar x) + \cdots
\right \} ,
\label{eq:qef} \\
\hspace*{-15mm} &&
\phi^{(g)} (x, Q) = x \bar x (x -  \bar x)
\left [ 16 B^{(q)}_2
\left ( \frac{\alpha_s (Q^2)}{\alpha_s (\mu_0^2)} \right )^{\frac{48}{81}}
+ 5 B^{(g)}_2
\left ( \frac{\alpha_s (Q^2)}{\alpha_s (\mu_0^2)} \right
)^{\frac{101}{81}}
\right ] + \cdots ,      
\label{eq:gef}
\end{eqnarray}
where $x$ and $\bar x = 1 - x$ are the energy fractions of two 
partons in the $\eta^\prime$-meson and $Q^2 > 0$ is the energy scale 
parameter. It is seen that in the limit $Q^2 \to \infty$ the quark 
wave-function~(\ref{eq:qef}) turns to its asymptotic form 
$\phi_{\rm as} (x) = 6 x \bar x$ (the same asymptotic behavior as the 
pion wave-function~\cite{BL} due to its quark-antiquark content), 
while the gluonic wave-function~(\ref{eq:gef}) vanishes in this limit,
$\phi^{(g)}_{\rm as} = 0$.  
The coefficients of the expansion of the wave-functions~(\ref{eq:qef}) 
and~(\ref{eq:gef}) are calculated by using  perturbation theory and 
include the effective QCD coupling $\alpha_s (Q^2)$, which in the 
next-to-leading logarithmic approximation is given by~\cite{PDG}
\begin{equation} 
\alpha_s (Q^2) = \frac{4 \pi}{\beta_0 \ln (Q^2 / \Lambda^2)} 
\left [ 1 - \frac{2 \beta_1}{\beta_0^2} \, 
\frac{\ln \ln Q^2 / \Lambda^2}{\ln Q^2 / \Lambda^2} \right ],
\label{eq:alpha-s} 
\end{equation}  
where $\beta_0 = 11 - 2 n_f / 3$, $\beta_1 = 51 - 19 n_f /3$, 
$\Lambda = \Lambda_{\rm QCD}$ is the QCD scale parameter,  
and $n_f$ is the number of quarks with masses less than the energy 
scale~$Q$. In the energy region $m_c^2 < Q^2 < m_b^2$, where 
$m_c$ and $m_b$ are the charm and  bottom quark masses, respectively, we
have used the central value of the dimensional parameter
corresponding to four active quark flavors,
$\Lambda^{(4)}_{\overline{MS}} = 280$~MeV~\cite{PDG}.

\section{The $\egg$ vertex function in the Brodsky-Lepage Approach}
\label{sec:FF-BL}

The invariant amplitude~${\cal M}$, corresponding to the effective $\egg$
vertex,
is the sum of the quark-antiquark~${\cal M}^{(q)}$ and gluonic~${\cal
M}^{(g)}$ components:
\begin{equation}
{\cal M} = {\cal M}^{(q)} + {\cal M}^{(g)} .
\label{eq:M-gen}
\end{equation}
The quark-antiquark~$F^{(q)}_{\egg}$ and gluonic~$F^{(g)}_{\egg}$
components of the
$\egg$ vertex function~$F_{\egg}$ entering in the invariant
amplitude are defined as follows~\cite{AP2000}:
\begin{equation}
{\cal M}^{(q, g)} \equiv - i \, F^{(q, g)}_{\egg}
(q_1^2, q_2^2, m_{\eta^\prime}^2) \,
\delta_{a b} \, \varepsilon^{\mu \nu \rho \sigma} \,
\varepsilon^{a*}_\mu \varepsilon^{b*}_\nu q_{1\rho} q_{2\sigma},
\label{eq:FF-def}
\end{equation}
where $\varepsilon^a_\mu$ and $\varepsilon^b_\nu$ are the polarization
vectors of the two gluons and $q_{1 \rho}$, $q_{2 \sigma}$ are their
four-momenta.

\begin{figure}[t!]
\includegraphics[bb=130 500 350 610,scale=1.0]{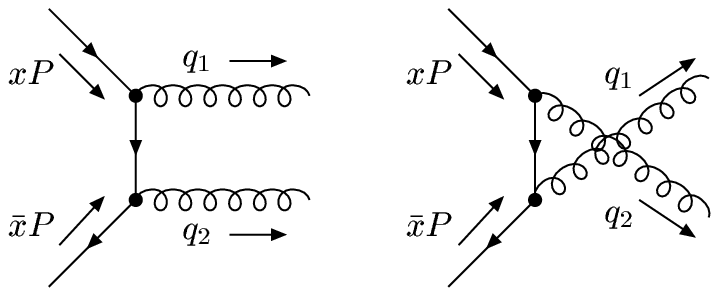}\\
\caption{Lowest order quark-antiquark contribution to the $\egg$ 
         vertex.}
\label{fig:quark-contrib}
\end{figure}

\begin{figure}[t!]
\includegraphics[bb=140 500 360 610,scale=1.0]{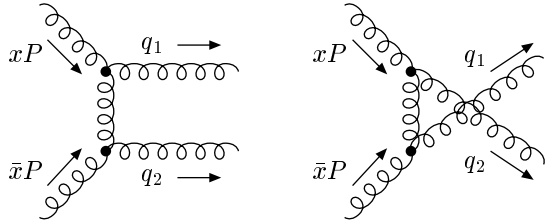}\\
\caption{Lowest order gluonic contribution to the $\egg$ vertex.}
\label{fig:gluon-contrib} 
\end{figure}

The diagrams depicting the quark-antiquark and gluonic contents of
the $\egg$ vertex are shown in Fig.~\ref{fig:quark-contrib}
and~\ref{fig:gluon-contrib}, respectively, 
and lead to the following vertex functions~\cite{AP2000}:
\begin{eqnarray}
F^{(q)}_{\egg} (q_1^2, q_2^2, m_{\eta^\prime}^2) & = &
4 \pi \alpha_s (Q^2) \, \frac{C}{2 N_c} \int\limits_0^1
dx \, \phi^{(q)} (x, Q)
\label{eq:FF-quark} \\ 
& \times &
\left [ \frac{1}
{x q_1^2 + \bar x q_2^2 - x \bar x m_{\eta^\prime}^2 + i \epsilon}
+ (x \leftrightarrow \bar x)
\right ] ~,
\nonumber \\ 
F^{(g)}_{\egg} (q_1^2, q_2^2, m_{\eta^\prime}^2) & = & 
\frac{4 \pi \alpha_s (Q^2)}{Q^2} \, \frac{C}{2} \int\limits_0^1
dx \, \phi^{(g)} (x, Q) 
\label{eq:FF-gluon} \\
& \times &
\left [
\frac{x q_1^2 + \bar x q_2^2 - (1 + x \bar x) m_{\eta^\prime}^2}
     {\bar x q_1^2 + x q_2^2 - x \bar x m_{\eta^\prime}^2 + i \epsilon}  
- (x \leftrightarrow \bar x) \right ] .
\nonumber 
\end{eqnarray} 
Note that the gluonic wave-function of the
$\eta^\prime$-meson~(\ref{eq:gef}) satisfies the antisymmetry condition
$\phi^{(g)} (x, Q) = - \phi^{(g)} (\bar x,
Q)$~\cite{Ohrndorf:1981uz,Shifman:1981dk,Baier:1981pm}.
It implies that if the relative sign in the brackets of
Eq.~(\ref{eq:FF-gluon}) {\it were} a ``$+$''
(as given in Eq.~(6) in Ref.~\cite{Muta}, and with which we differ) the
gluonic contribution
to the $\egg$ form factor would vanish identically.

\begin{figure}[t!]
\includegraphics[bb=120 445 450 710,scale=0.75]{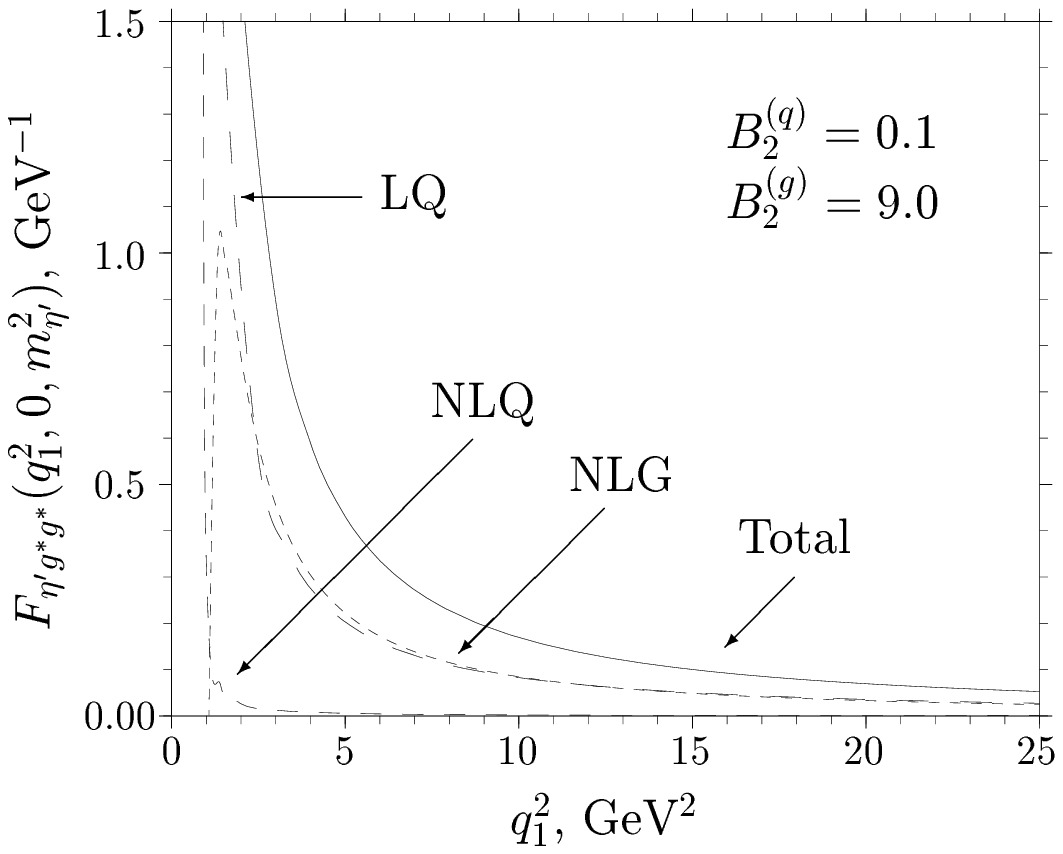}
\includegraphics[bb=120 445 450 710,scale=0.75]{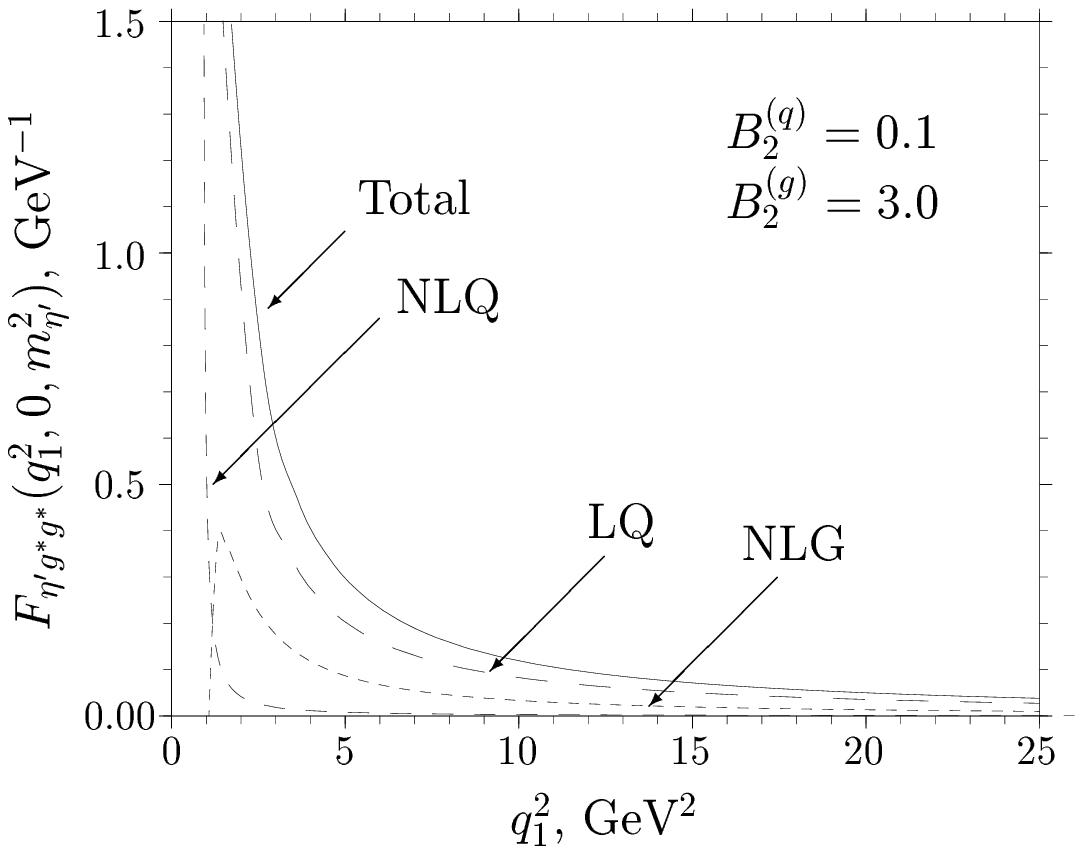}
\caption{The $\eta^\prime g^* g$ vertex
         $F_{\egg} (q_1^2,0,m_{\eta^\prime}^2)$ as a function
         of $q_1^2$ with $B^{(q)}_2 = 0.1$ and two values of~$B^{(g)}_2$:
         $B^{(g)}_2 = 9.0$ (upper plot) and $B^{(g)}_2 = 3.0$ (lower
         plot) in the Brodsky-Lepage approach.
         The dashed curves are the leading (LQ), next-to-leading
         quark-antiquark (NLQ), and gluonic (NLG) components, and the
         solid curve is the sum.}
\label{fig:form-fac}
\end{figure}

The quark and gluonic wave-functions contain both free
($B^{(q)}_n$, $B^{(g)}_n$) and constrained parameters, with the latter
depending on the anomalous dimensions.
The free parameters can be fitted from the experimental data,   
for example, from the $\eta^\prime \gamma^* \gamma$ transition form
factor. We take the following restrictions on the first correction to the
leading order quark-antiquark wave-function~: $|B^{(q)}_2| < 0.1$ and
$|\rho^{(g)}_2 \, B^{(g)}_2| < 0.1$ in order to keep $\phi^{(q)} (x, Q)$
close to its asymptotic value: $\phi_{\rm as} (x) = 6 x \bar x$, in
agreement with the experimental data on the $\eta^\prime \gamma^* \gamma$
transition form factor~\cite{eta'-gamma}. Taking into account the
anomalous dimension-dependent quantity  
$\rho^{(g)}_2 \simeq - 1/90$ from Ref.~\cite{AP2000}
we get\footnote{This differs considerably
from the values used in~\cite{Muta}.} $|B^{(g)}_2| < 9.0$.  
Below, we shall present the $\eta^\prime g^* g^*$ vertex function
for the maximum allowed values of the non-perturbative parameters,
i.e., $|B^{(q)}_2| = 0.1$ and $|B^{(g)}_2| = 9.0$. We note that
there is not much sensitivity to the variation  of the
parameter~$B^{(q)}_2$ on the overall $\eta^\prime g^* g^*$ vertex
function . Hence, we fix this parameter to its maximum
allowed value $|B^{(q)}_2| = 0.1$. However, there is considerable  
sensitivity to the variation of the parameter~$B^{(g)}_2$. To show this
we shall take $|B^{(g)}_2| = 9.0$  and $|B^{(g)}_2| = 3.0$. The resulting
theoretical dispersion between the two cases ($|B^{(g)}_2| = 9.0$ vs.
$|B^{(g)}_2| = 3.0$) can be seen in Fig.~\ref{fig:form-fac}, where we show
the $\eggr$ vertex function for an on-shell gluon ($q_2^2=0$) with the
other having a time-like virtuality ($q_1^2 > 0$) in the Brodsky-Lepage
approach. The various leading order and next-to-leading order components
and the sum contributing to $F_{\eta^\prime g^*
g}(q_1^2,0,m_{\eta^\prime}^2)$ are displayed individually.
The gluonic contribution
(called NLG) for the maximum values of the free parameters is
comparable to the leading quark contribution (called LQ), as shown in
the upper plot in Fig.~\ref{fig:form-fac}, increasing the
$\eta^\prime g^* g$ vertex by almost a factor 2 as compared to the
case when only the quark content of the $\eta^\prime$-meson is assumed.
This may be considered as the maximum gluonic content
of the  $\eta^\prime$-meson allowed by current data. Even in the 
more realistic case with $B^{(q)}_2 = 0.1$ and $B^{(g)}_2 = 3.0$, we
see from the lower plot in Fig.~\ref{fig:form-fac} that the
gluonic contribution  is not small, and also in this case it enhances
the value of the total $\eta^\prime g^* g$ vertex function (the solid
curve in Fig.~\ref{fig:form-fac}) at the level of few tens percent.


\section{The $\egg$ vertex function in the mHSA Approach}
\label{sec:FF-mHSA}

In the Brodsky-Lepage approach to form factors, the transverse momentum
dependence of the partons in the mesons is neglected 
as the hard scattering (perturbative) approach is applicable only when the
virtualities of the external particles are much larger than the typical
value of the parton transverse momenta~${\bf k}_{\perp i}$. 
Including the transverse momenta of the partons in the meson, the
perturbative expansion of the transition form factor encounters large
logarithms of the form $\ln (Q^2 / {\bf k}_{\perp i}^2)$, and it becomes
mandatory to sum the  multiple-gluon emissions. The formalism for the 
soft and collinear gluon resummation was introduced by Collins and
Soper~\cite{CS} and by Collins, Soper and Sterman~\cite{CSS}. Such gluon
emissions give rise to powers of double logarithms in each order of
perturbation theory and their contribution exponentiates into the 
Sudakov function~\cite{Sudakov}. The Sudakov exponents are known both
for the quark-antiquark case, from the Drell-Yan (DY) and the deep
inelastic scattering (DIS) processes, and for gluons from the gluon fusion
into $2\gamma$, gauge, or Higgs boson final states.
This formalism is suitable 
for the description of the hadronic wave-functions and the hadronic
form factors, such as the electromagnetic and transition form factors of
the 
pion~\cite{BS89,JKR96,JK93,Sudakov-gen,Gousset}, and has also been 
employed in calculating the $\eta^\prime g^{*}g^{*}$ vertex in
Ref.~\cite{AP2000}.

In the modified Hard Scattering Approach (mHSA)~\cite{BS89}, 
we take the $\eta^\prime$-meson wave-function in a form
similar to the pion wave-function~\cite{JK93,JKR96}:
\begin{equation}
\hat \Psi^{(p)} (x, Q, {\bf b}) =
\frac{2 \pi C}{\sqrt{2 N_c}} \,
\phi^{(p)} (x, Q) \,
\exp \left [ - \frac{x \bar x b^2}{4 a^2} \right ] \, 
S^{(p)} (x, Q, b) ,
\label{eq:WF-gen} 
\end{equation}
where $p = q$ for the quark and antiquark case, and $p = g$ for
gluons;
the constant~$C$ is already defined in Eq.~(\ref{eq:C-const}), ${\bf b}$
is the separation between the $\eta^\prime$-meson constituents
in the transverse configuration space, with $b=\vert {\bf b}\vert$ often
called the impact
parameter, and $\phi^{(q)}$ and $\phi^{(g)}$ have the form presented
in Eqs.~(\ref{eq:qef}) and~(\ref{eq:gef}), respectively.
The transverse size parameter~$a$ can be determined from the average 
transverse momentum of the $\eta^\prime$-meson. For the numerical
analysis, the value $a^{-1} = 0.861$~GeV is used, following from the
analysis of the form factor involving the
$\pi$-meson~\cite{JKR96,JK93}.
The soft-gluon emission from the quark, antiquark and gluons in
the $\eta^\prime$-meson can be taken into account by including the
QCD Sudakov factors $S^{(p)} (x, Q, b)$, the details of which can
be found in Ref.~\cite{AP2000}. 

\begin{figure}[t!]
\includegraphics[bb=125 445 455 715,scale=0.85]{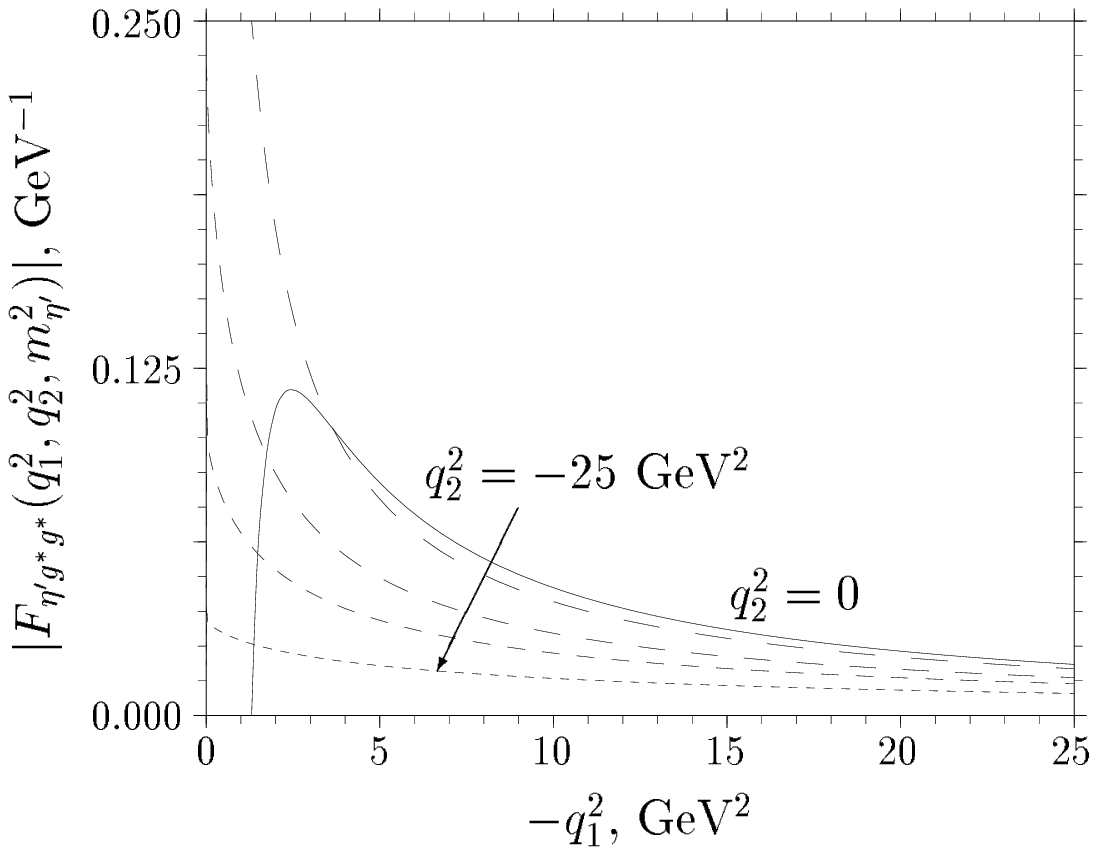}
\includegraphics[bb=125 445 455 715,scale=0.85]{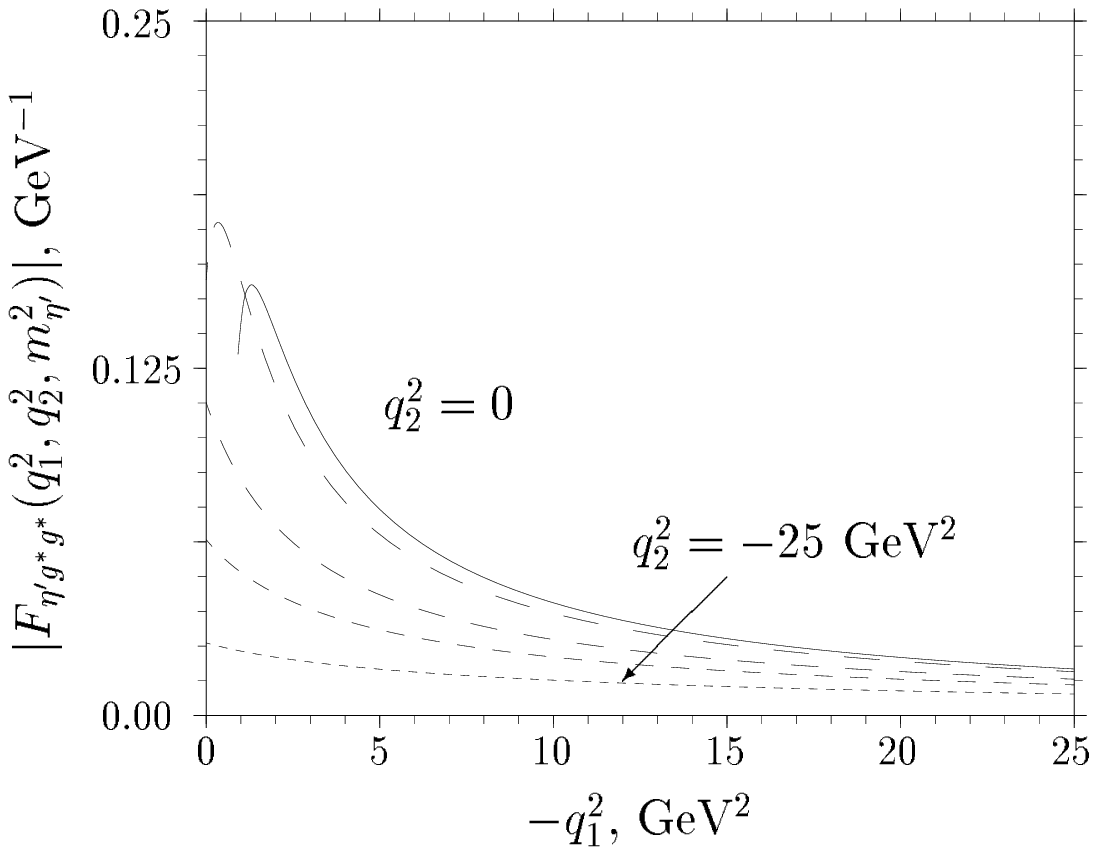}
\caption{ The $\egg$ vertex function $|F_{\eta^\prime g^* g^*} 
         (q_1^2, q_2^2, m_{\eta^\prime}^2)|$ for space-like
         gluon virtualities
         in the Brodsky-Lepage approach (upper figure) and
         in the mHSA formalism (lower figure), with $B^{(q)}_2 = 0.1$
         and $B^{(g)}_2 = 3.0$.  The legends are as follows:
         $q_2^2 = 0$ (solid curve), $q_2^2 = -1$~GeV$^2$ (long-dashed
         curve), $q_2^2 = -5$~GeV$^2$ (medium-dashed curve),
         $q_2^2 = -10$~GeV$^2$ (short-dashed curve), and
         $q_2^2 = -25$~GeV$^2$ (dotted curve).}
\label{fig:ff-bljk-mp}
\end{figure}

\begin{figure}[t!]
\includegraphics[bb=125 445 455 715,scale=0.85]{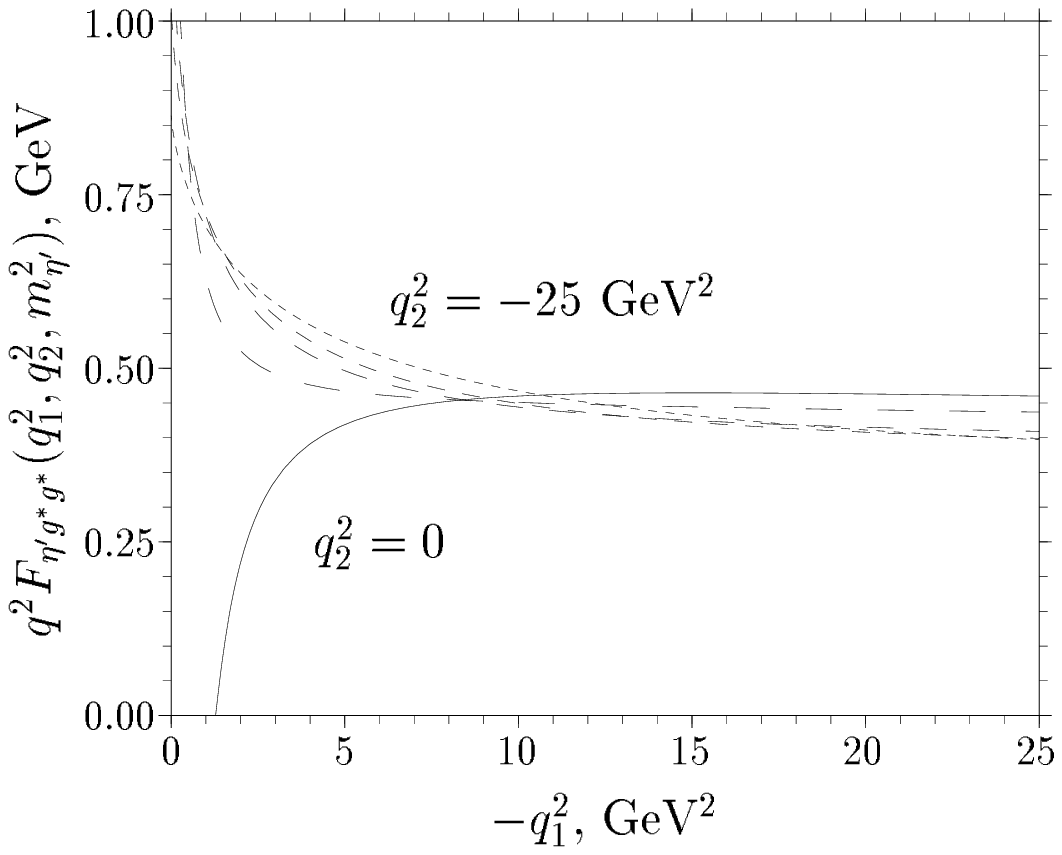}
\includegraphics[bb=125 445 455 715,scale=0.85]{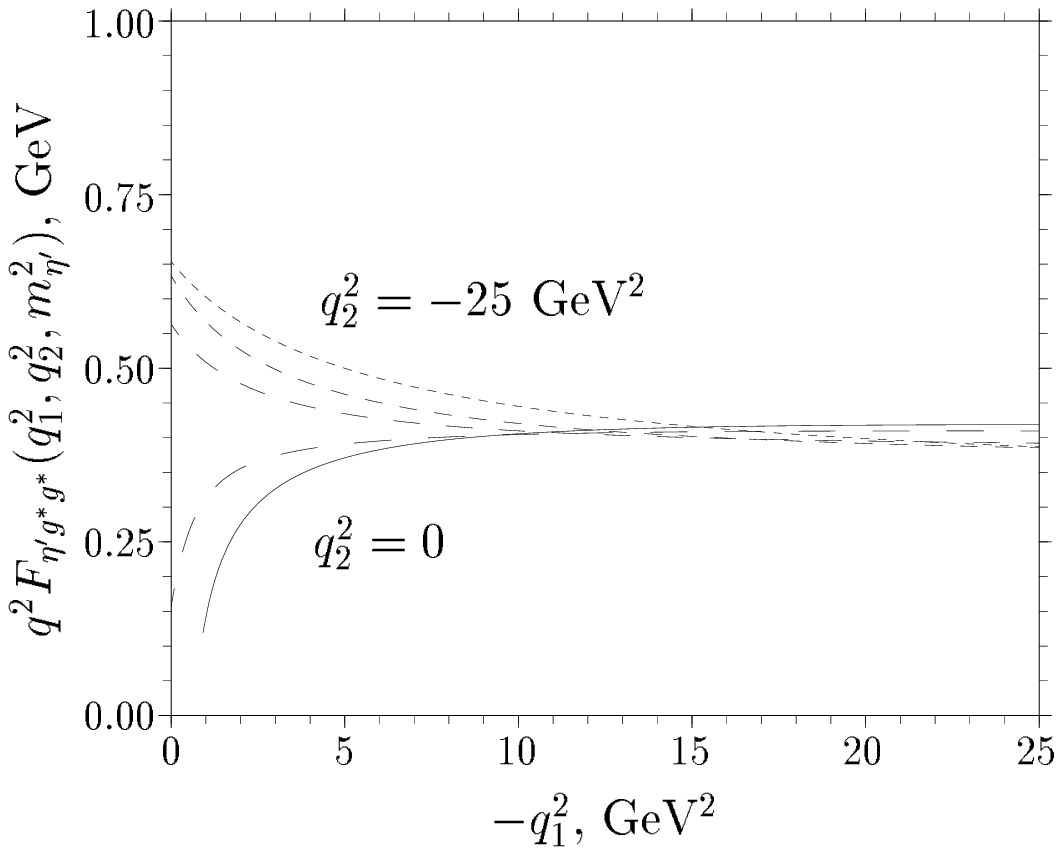}
\caption{The $\egg$ vertex function $q^2 F_{\eta^\prime g^* g^*}
         (q_1^2, q_2^2, m_{\eta^\prime}^2)$ for space-like
         gluon virtualities
         in the Brodsky-Lepage approach (upper figure) and
         in the mHSA formalism (lower figure), with $B^{(q)}_2 = 0.1$
         and $B^{(g)}_2 = 3.0$. Legends are the same as in
         Fig.~\protect\ref{fig:ff-bljk-mp}.}
\label{fig:ff-bljk-mq}
\end{figure}

 In the space-like region of the gluon virtualities, the quark and gluonic
vertex functions are \cite{AP2000}:  
\begin{eqnarray}
F^{(q)}_{\egg} (Q, \omega, \eta) & = & - 4 \pi \alpha_s (Q^2) \, 
\frac{C}{N_c \Lambda^2} \int\limits_0^1 dx \, \phi^{(q)} (x, Q)
\nonumber \\
& \times &
\int\limits_0^1 db_\Lambda \, b_\Lambda \,
\exp \left [ - \frac{x \bar x}{4 a^2 \Lambda^2} \, b_\Lambda^2 \right ] \,
S^{(q)} ( x, Q, b) \, K_0^{(+)} (x, b_\Lambda Q_\Lambda) ,
\label{eq:FF-q-mHSA-m} \\
F^{(g)}_{\egg} (Q, \omega, \eta) & = & - 4 \pi \alpha_s (Q^2) \, 
\frac{C}{\Lambda^2} \int\limits_0^1 dx \, \phi^{(g)} (x, Q) \, 
\int\limits_0^1 db_\Lambda \, b_\Lambda \,
\exp \left [ - \frac{x \bar x}{4 a^2 \Lambda^2} \, b_\Lambda^2 \right ]
S^{(g)} (x, Q, b) \,
\nonumber \\
& \times & 
\left [ |\eta| \, K_0^{(-)} (x, b_\Lambda Q_\Lambda) -
(x - \bar x) \omega \, K_0^{(+)} (x, b_\Lambda Q_\Lambda) \right ] ~.
\label{eq:FF-g-mHSA-m} 
\end{eqnarray} 
Here,
\begin{equation}
K^{(\pm)}_0 (x, b_\Lambda Q_\Lambda) = 
\frac{1}{2} \left [
K_0 \left ( b_\Lambda Q_\Lambda \lambda_+ (x, \omega, \eta) \right )
\pm
K_0 \left ( b_\Lambda Q_\Lambda \lambda_+ (\bar x, \omega, \eta) \right )
\right ] ~,
\nonumber 
\end{equation}
where the various dimensionless parameters are defined as: $b_\Lambda = b
\Lambda$, $Q_\Lambda = Q / \Lambda$, 
$|\eta| = m_{\eta^\prime}^2 / Q^2$, $\omega = (q_1^2 - q_2^2)/Q^2$, 
$K_0 (z)$ is the modified Bessel function, and 
\begin{eqnarray}
\lambda_\pm^2 (x, \omega, \eta) = \frac{1}{2} \,
\left [ 1 + \omega (x - \bar x) \pm 2 x \bar x |\eta| \right ]. 
\label{eq:Lambda-func}
\end{eqnarray}
The vertex function $|F_{\eta^\prime g^* g^*} (q_1^2, q_2^2,
m_{\eta^\prime}^2)|$ calculated in the space-like region is
plotted in Fig.~\ref{fig:ff-bljk-mp} as a function of $q_1^2$ for given
values of $q_2^2$ for the Brodsky-Lepage case (upper figure) and in the
mHSA formalism (lower figure). The function $q^2 F_{\eta^\prime g^* g^*}
(q_1^2,q_2^2, m_{\eta^\prime}^2)$ in the two approaches is shown in
Fig.~\ref{fig:ff-bljk-mq}. We note the improved perturbative behaviour
of the vertex functions for smaller virtualities in the mHSA formalism,
while for larger virtualities the two approaches yield very similar
results. 

\begin{figure}[t!]
\includegraphics[bb=135 360 450 665,scale=0.65]{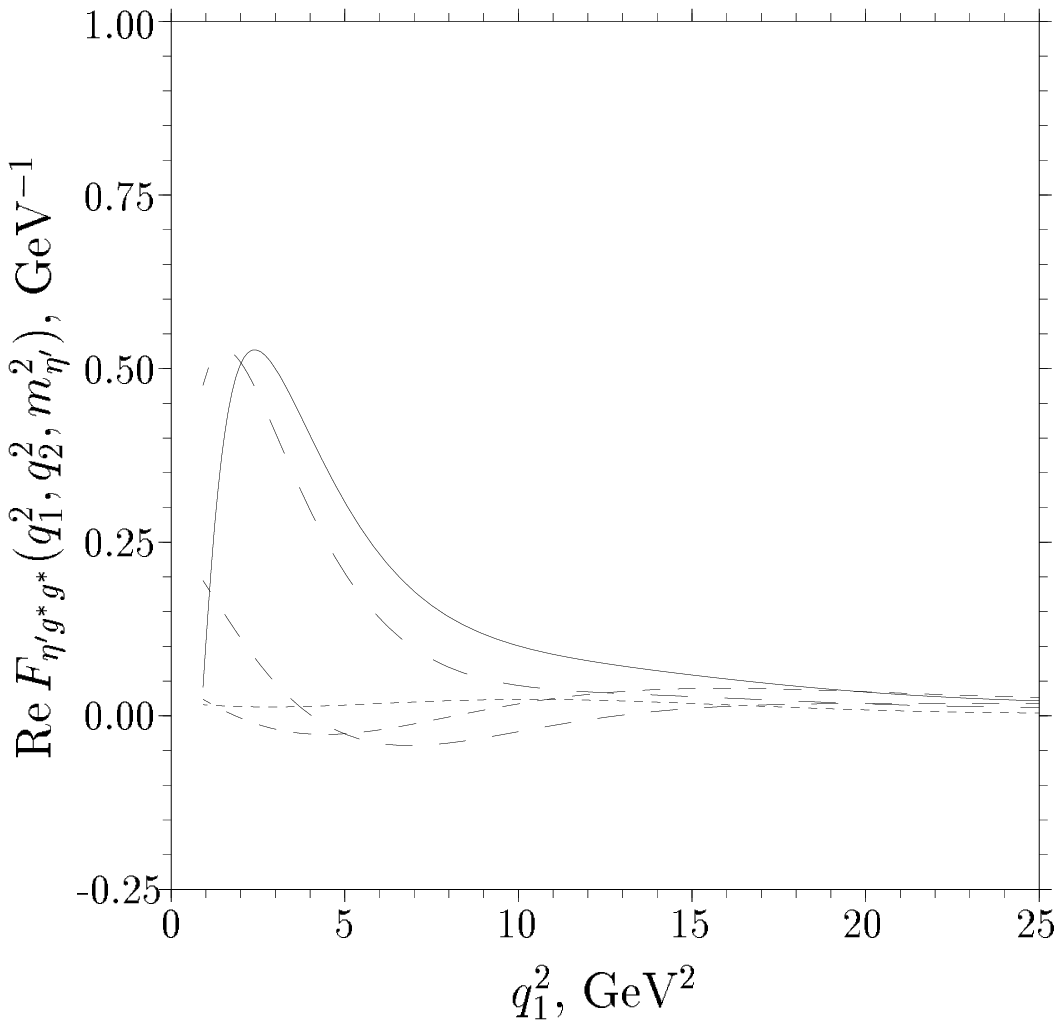}
\includegraphics[bb=135 360 450 665,scale=0.65]{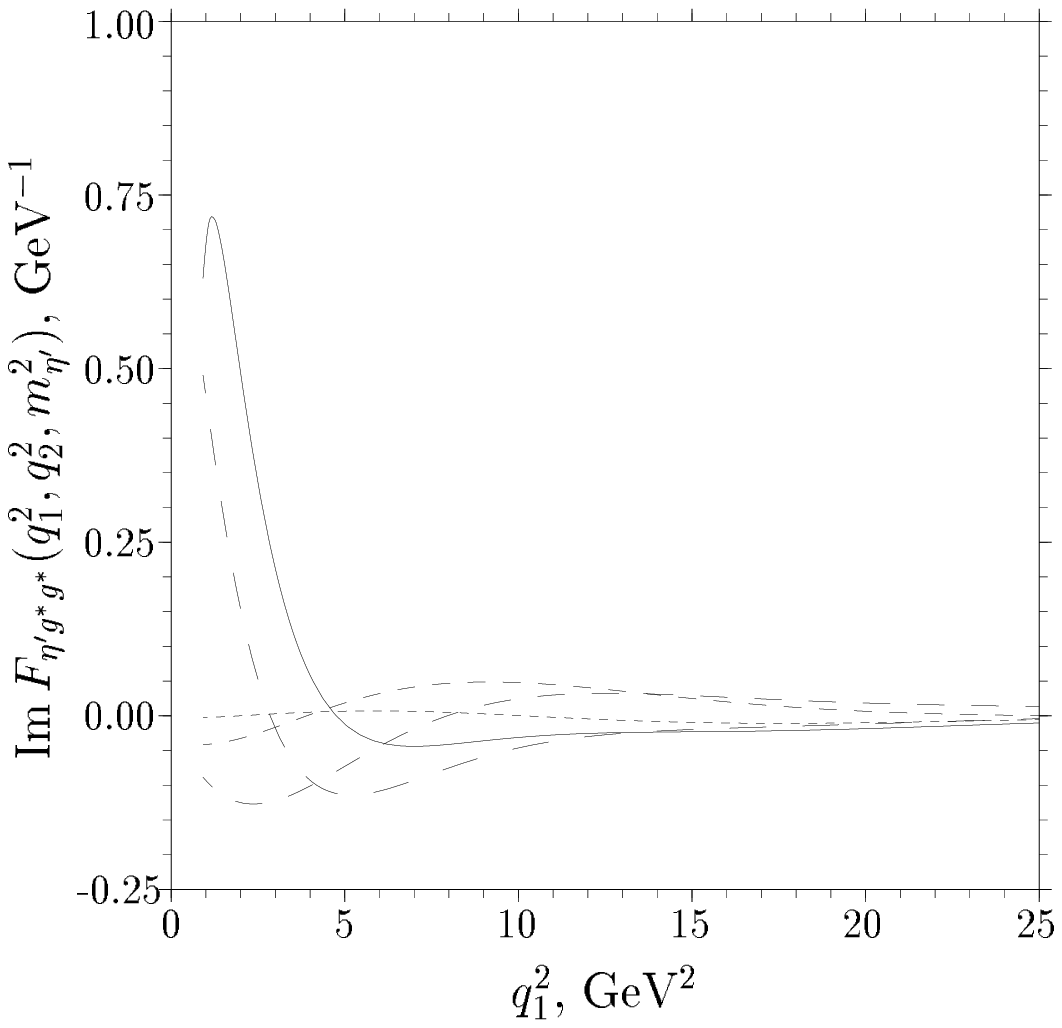}
\includegraphics[bb=135 400 450 665,scale=0.65]{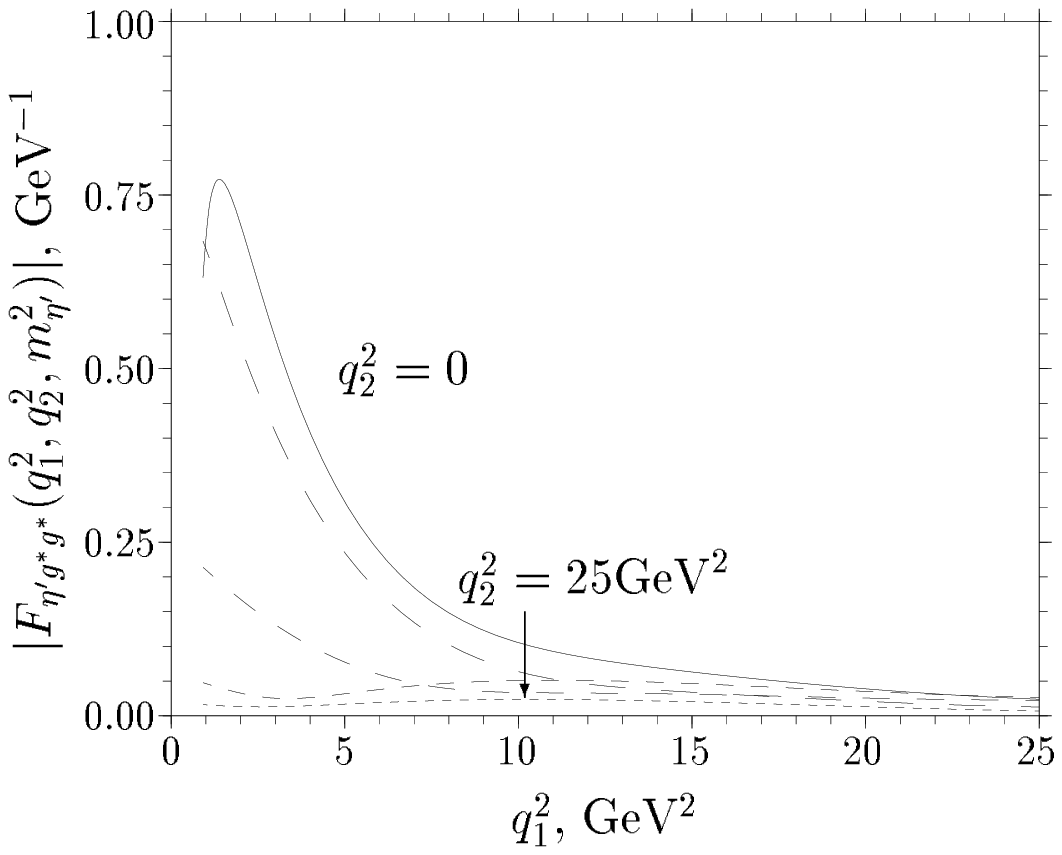}
\includegraphics[bb=130 450 450 710,scale=0.65]{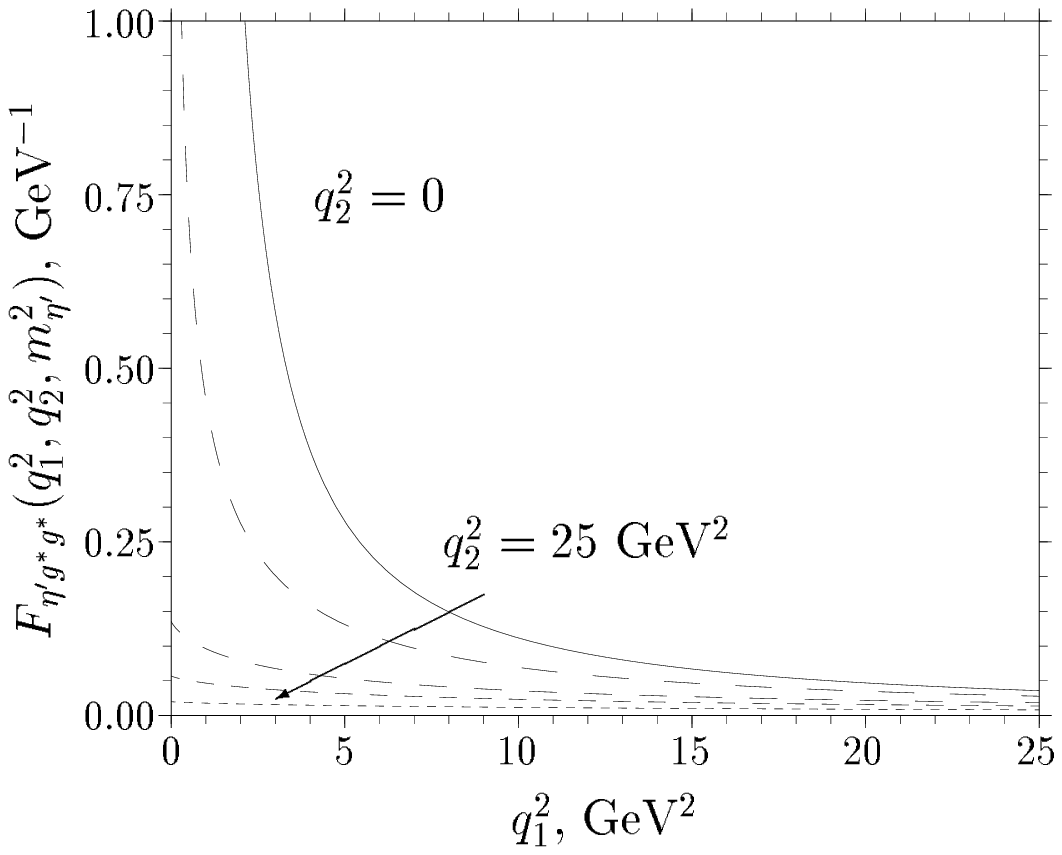}
\caption{The real and imaginary parts of the $\egg$ vertex
         $F_{\eta^\prime g^* g^*} (q_1^2, q_2^2, m_{\eta^\prime}^2)$
         in the mHSA formalism (top two figures), and its magnitude
 $\vert F_{\eta^\prime g^* g^*} (q_1^2, q_2^2, m_{\eta^\prime}^2)\vert$
         (bottom left figure) with time-like gluon virtualities,
         $B^{(q)}_2 = 0.1$ and $B^{(g)}_2 = 3.0$.  
         The bottom right hand figure shows the form factor in the
         Brodsky-Lepage approach. The legends are as follows:
         $q_2^2 = 0$ (solid curve), $q_2^2 = 1$~GeV$^2$ (long-dashed
         curve), $q_2^2 = 5$~GeV$^2$ (medium-dashed curve),
         $q_2^2 = 10$~GeV$^2$ (short-dashed curve), and
         $q_2^2 = 25$~GeV$^2$ (dotted curve).} 
\label{fig:ff-jk-p}
\end{figure}

The transition from the space-like region of gluon virtualities 
to the time-like one for the $\egg$ vertex function is discussed 
in Ref.~\cite{AP2000}. The final result for the quark and gluonic
contributions in the time-like region is: 
\begin{eqnarray}
F^{(q)}_{\egg} (Q, \omega, \eta) & = & 
4 \pi \alpha_s (Q^2) \, \frac{i \pi C}{2 N_c \Lambda^2}
\int\limits_0^1 dx \, \phi^{(q)} (x, Q)
\nonumber \\
& \times &
\int\limits_0^1 db_\Lambda \, b_\Lambda \,
\exp \left [ - \frac{x \bar x}{4 a^2 \Lambda^2} \, b_\Lambda^2 \right ]
S^{(q)} ( x, Q, b) \, H_0^{(+)} (x, b_\Lambda Q_\Lambda) ,
\label{eq:FF-q-mHSA-p} \\
F^{(g)}_{\egg} (Q, \omega, \eta) & = & 
4 \pi \alpha_s (Q^2) \, \frac{i \pi C}{2 \Lambda^2}
\int\limits_0^1  dx \, \phi^{(g)} (x, Q) \, 
\int\limits_0^1 db_\Lambda \, b_\Lambda \,
\exp \left [
- \frac{x \bar x}{4 a^2 \Lambda^2} \, b_\Lambda^2 \right ] \, 
S^{(g)} ( x, Q, b ) \,
\nonumber \\
& \times & 
\left [
\eta \, H_0^{(-)} (x, b_\Lambda Q_\Lambda)
+ (x - \bar x) \omega \, H_0^{(+)} (x, b_\Lambda Q_\Lambda)
\right ] ~,
\label{eq:FF-g-mHSA-p}
\end{eqnarray}
where
\begin{equation}
H_0^{(\pm)} (x, b_\Lambda Q_\Lambda) =  
\frac{1}{2} \bigg [H^{(2)}_0   
\left (b_\Lambda Q_\Lambda \lambda_- (x, \omega, \eta ) \right )
\pm  H^{(2)}_0
\left ( b_\Lambda Q_\Lambda \lambda_- (\bar x, \omega, \eta) \right )
\bigg ]. 
\nonumber 
\end{equation}
Here, $H^{(2)}_0 (z)$ is the second Hankel function \cite{Bateman}.
It is interesting to note that due to the $i\varepsilon$ prescription 
of the propagators in the hard scattering part of the $\egg$ vertex
function, an imaginary part is generated. This
can be seen in Fig.~\ref{fig:ff-jk-p}, where the upper two figures  show
the real and imaginary parts of the vertex function  
$F_{\eta^\prime g^* g^*} (q_1^2, q_2^2, m_{\eta^\prime}^2)$
as a function of $q_1^2$ for the indicated values of $q_2^2$, calculated
in the mHSA approach. In the lower two figures, we  compare the
magnitude of the vertex function in this approach with the one in the
Brodsky-Lepage approach. A comparison shows that there is a marked
difference between the vertex functions calculated in the Brodsky-Lepage
and the mHSA approaches, pertaining to the absence of the singularity at
$Q^2 = m_{\eta^\prime}^2$~\cite{AP2000} in the latter case. 

\section{Interpolating Expressions for the $\egg$ Vertex Functions} 
\label{sec:asymptotics} 

For the applications in various decay and production processes it is
useful to find an approximate expression for the vertex function
which is simple and can be used over a
large domain of the gluon virtualities. We recall that
Brodsky and Lepage~\cite{BL81} presented an
approximate form for the $\pi-\gamma$ transition form factor
which interpolates between the PCAC value and the QCD prediction
in the large~$Q^2$ region. Subsequently, in Ref.~\cite{FK98-2}, this
form was extended to the case of the $\eta^\prime-\gamma$ transition form
factor. Very much along the same lines, a similar expression can be
written for the $\eta^\prime g^* g^*$ transition form factor~\cite{AP2000}:
\begin{equation}
F^{\rm BL}_{\egg} (q^2, \omega) = 4 \pi \alpha_s (Q^2) \,
\frac{2 \sqrt{3} f_\pi D (q^2, \omega)}
     {\sqrt{3} \, q^2 - 8 \pi^2 f_\pi^2 D (q^2, \omega)} ,
\label{eq:BL-appr}
\end{equation}
where the largest energy scale parameter $Q^2 = q^2$ for the time-like
total gluon virtuality and $Q^2 = - q^2$ for the space-like one.
The function $D (q^2, \omega)$ is defined as follows \cite{AP2000}:
\begin{eqnarray}
D (q^2, \omega) & = & f_0 (\omega) + \frac{q^2}{Q^2} \,
\left [ 16 B^{(q)}_2
\left ( \frac{\alpha_s (Q^2)}{\alpha_s (\mu_0^2)} \right )^{\frac{48}{81}}
\! \! \!
+ 5 B^{(g)}_2 \left ( \frac{\alpha_s (Q^2)}{\alpha_s (\mu_0^2)}
\right )^{\frac{101}{81}}
\right ] g_2 (\omega) , 
\label{eq:D-function} \\ 
f_0 (\omega) & = & \frac{1}{\omega^2}
\left [
1 - \frac{1 - \omega^2}{2 \omega} \ln
\left |
\frac{1 + \omega}{1 - \omega}
\right |
\right ] ,
\qquad
g_2 (\omega) = \frac{3 f_0 (\omega) - 2}{6\omega} , 
\nonumber
\end{eqnarray}
where $\omega$ is the asymmetry parameter having values in
the interval $0 \le \omega \le 1$. The asymptotic functions for the
quark-antiquark~$f_0 (\omega)$ and gluon~$g_2 (\omega)$ cases are
derived in Ref.~\cite{AP2000} and obey the bounds
$2/3 \le f_0 (\omega) \le 1$ and  $0 \le g_2 (\omega) \le 1/6$. 
We remark that the difference between the asymptotic behaviour 
of the vertex functions in the Brodsky-Lepage and mHSA
approaches is small~\cite{AP2000}. 

The above expression reproduces both the anomaly value and
the large~$Q^2$ asymptotics of the vertex functions:
\begin{eqnarray}
\left . F_{\egg} (Q^2, \omega) \right |_{Q^2 \to 0} & = &
4 \pi \alpha_s (m_{\eta^\prime}^2) \, \frac{\sqrt 3}{4 \pi^2 f_\pi} ,
\label{eq:asymp-0} \\
\left . F_{\egg} (Q^2, \omega) \right |_{Q^2 \to \infty}
& = & 4 \pi \alpha_s (Q^2) \, \frac{2 f_\pi D (q^2, \omega)}{q^2} .
\label{eq:asymp-inf}
\end{eqnarray}
Presented in Eq.~(\ref{eq:BL-appr}), the interpolating function is a
smooth function in the space-like region of the gluon virtualities but has
a pole at $q^2 = 8 \pi^2 f_\pi^2 D(q^2, \omega) / \sqrt 3$ in the
time-like region. A similar behavior for the form factor in the
time-like region was obtained by Kagan and Petrov~\cite{TFF-1} as the
result of the evaluation of the triangle diagram.

The mHSA approach naturally removes the unphysical singularity from
the vertex function in the time-like region of the gluons virtualities
but the vertex function gets an imaginary part. In this case the real
part of the approximate formula interpolates between the anomaly value and
the large~$Q^2$ asymptotics, while the imaginary part goes to zero
as $Q^2 \to 0$. An approximate interpolating form in the time-like
region of the gluon virtualities is given in Ref.~\cite{AP2000} to which
we refer for further details. 

\section{Summary}
\label{sec:summary}

We have studied the $\egg$ vertex function
$F_{\egg} (q_1^2, q_1^2, m_{\eta^\prime}^2)$ in perturbation theory
for the most general case when both gluons are virtual. 
The evolution equations involving the eigenfunctions of the
quark-antiquark and gluonic components of the $\eta^\prime$-meson wave 
function are solved, and the input
parameters in the wave function 
are determined using data on the electromagnetic transition form factor
of the
$\eta^\prime$-meson. 
It is shown that within the allowed variation of the  
parameters $B^{(q)}_2$ and $B^{(g)}_2$ of the $\eta^\prime$-meson
wave-function, the gluonic contribution is not small. For extremal values
of the parameters allowed by data, $B^{(q)}_2 = 0.1$ and $B^{(g)}_2 =
9.0$, the gluonic correction is found to be comparable to the quark
contribution. But, even for a lower value $B^{(g)}_2 = 3.0$, the gluonic
contribution is present at the level of few tens percent. Our work 
corrects and extends the existing results on the $\eta^\prime g^* g$ 
vertex function,
reported earlier in Ref.~\cite{Muta} for the case of one virtual
gluon in the time-like region.  

We find that the Brodsky-Lepage approach leads to the
appearance of a singularity in the region of the $\eta^\prime$-meson
mass for time-like gluon virtualities. This reflects the observation made
earlier in the literature that  
the Brodsky-Lepage approach to exclusive form factors is valid only in the
asymptotic region. In conformity with this, we find that for the
total gluon virtuality $|q^2| = | q_1^2 + q_2^2 | \gg m_{\eta^\prime}^2$, 
the $\egg$ vertex function has the usual asymptotic behaviour: 
$F_{\egg} \sim 1 / q^2$, anticipated for the pseudoscalar meson
transition form factors.

In the mHSA approach, where the transverse momentum dependence of the
hard scattering amplitude as well as the transverse momentum distribution
and the soft-gluon emission (the Sudakov factor) in the
$\eta^\prime$-meson wave-function are taken into account,  
the mentioned singularity in the Brodsky-Lepage-approach at $q^2 =
m_{\eta^\prime}^2$
disappears. Also, the validity of the perturbative approach is extended
to smaller virtualities in the mHSA formalism, though asymptotically the
two approaches yield very similar results for the space-like
gluon virtualities.
In the time-like region, in the mHSA formalism, 
we find that due to the $i \varepsilon$ prescription of the propagators,
the $\egg$ vertex function obtains an additional phase factor in comparison
with the space-like expression. Thus, in applications where the $\egg$
vertex appears with off-shell gluons in the time-like region, this
perturbative phase should be included. 

An approximate expression for the $\egg$ vertex function is presented 
for the case when both gluons are off mass shell and have 
space-like virtualities. This expression
interpolates between the anomaly value of the $\egg$ vertex and the
asymptotic QCD prediction in the large~$Q^2$ region.
The results summarized here have obvious applications in
rare $B$-meson decays $B \to \eta^\prime K$ and $B \to \eta^\prime X_s$,
and in a number of other radiative and hard processes involving the
vertex $\egg$.

\begin{acknowledgments}
A.P. would like to thank the DESY theory group for its hospitality in
Hamburg where the major part of this work was done. This talk was also
presented by A.P. at the DESY Theory Workshop ''CP Violation and Rare
Processes: Standard Model and Beyond'', Hamburg, September 26 - 29, 2000.
\end{acknowledgments}

%

%
%
\end{document}